# Dirac phonons in two-dimensional materials

Jialin Gong[a,1], Jianhua Wang[a,1], Hongkuan Yuan[a], Zeying Zhang[b,*], Xiaotian Wang[a,c,*]

[a]*School of Physical Science and Technology, Southwest University, Chongqing 400715, China*
[b]*College of Mathematics and Physics, Beijing University of Chemical Technology, Beijing 100029, China;*
[c]*Institute for Superconducting and Electronic Materials (ISEM), University of Wollongong, Wollongong 2500, Australia*
[1]*J.G. and J. W. contributed equally to this manuscript.*
Email addresses:
zzy@mail.buct.edu.cn;
xiaotianwang@swu.edu.cn

Phonons are an ideal platform for realizing stable spinless two-dimensional (2D) Dirac points because they have a bosonic nature and hard-to-break time-reversal symmetry. It should be noted that the twofold degenerate nodal points in the phonon dispersions of almost all reported 2D materials are misclassified as "Dirac points" owing to a historical issue. The correct name for these twofold degenerate nodal points should be "Weyl" because 2D phononic systems are essentially spinless and because each twofold degenerate point is described by a Weyl model in two dimensions. To date, there have been *no reports of fourfold degenerate Dirac point phonons in 2D materials.* In this study, we searched through the entire 80 layer groups (LGs) and discovered that Dirac phonons can be realized in 7 of the 80 LGs. Moreover, the Dirac points in the phonon dispersions of 2D materials can be divided into essential and accidental degenerate points, which appear at high-symmetry points and on high-symmetry lines, respectively. Guided by symmetry analysis, we identified the presence of Dirac phonons in several 2D material candidates with six LGs. This letter offers a method for identifying Dirac phonons in 2D and proposes 2D material candidates for realizing Dirac phonons.

The discovery of topological quantum states [1-4] is one of the most promising advancements in condensed matter physics. Currently, the study of topological quantum states is no longer limited to only electron-related systems [6-20] but has been widely extended to include bosonic systems [21-30]. Among the different types of bosonic systems, phonons [31-39], which are induced by atomic vibrations at THz frequency, are a perfect platform for realizing topological quantum states because of their unique advantages and possible applications. Notably, unlike that of electronic bands, the entire frequency range of phonon bands is relevant for experimental detection because phonons are not constrained by the Pauli exclusion principle and Fermi surface. Additionally, topological phonons [31,32,40] play critical roles in thermal transports, electron-phonon coupling, or multiphonon processes.

In the past five years, various types of symmetry-enforced topological phonons [40-60], including conventional and unconventional Weyl point phonons, triple point phonons, Dirac point phonons, nodal line phonons, and nodal surface phonons, have been discovered in three-dimensional (3D) realistic materials. Some of them have also been verified through experiments. Two-dimensional (2D) materials have a lower symmetry than 3D materials. Therefore, 2D materials with less symmetrical constraints may more intuitively display the clean characteristics of topological phonons. Unfortunately, topological phonons have only been explored in only a few 2D materials [61-64]. Systematical research into 2D topological phonons and their related material realizations is highly required. In 2022, Yu *et al.* [62] discovered the appearance of 2D twofold degenerate quadratic nodal point phonons and presented a guideline for studying the quadratic nodal point in 2D phononic systems via symmetry analysis. In 2020, Li *et al.* [63] proposed that 2D graphene hosts four types of "Dirac phonons" and a nodal ring phonon in its phonon spectrum. In 2018, Jin *et al.* [64] predicted the appearance of Dirac phonons in 2D hexagonal lattices using first-principles calculations. *Note that the linearly dispersed twofold phonon band crossing points in 2D graphene and 2D hexagonal lattices are misclassified as "Dirac points" because of their historical use in early graphene research* [65,66]. The correct name for these twofold degenerate nodal points should be "Weyl" because 2D phononic systems are essentially spinless and because each twofold degenerate band crossing point at K or $K_0$ in 2D graphene and 2D hexagonal lattices is characterized by a 2D Weyl Hamiltonian with a defined chirality.

*To the best of our knowledge, the study of Dirac phonons in 2D materials is still rather primitive and has not been undertaken by other researchers.* In this study, using 2D spinless phononic systems as targets, we searched through the entire 80 layer groups (LGs) with time-reversal symmetry $\mathcal{T}$

and identified 7 LG candidates that host Dirac points based on theoretical analysis (see Table I). Moreover, the Dirac points in the phonon dispersions of 2D materials can be divided into essential and accidental degenerate points, which appear at high-symmetry points (HSPs) and on high-symmetry lines (HSLs), respectively. It is well known that material realization is a prerequisite for studying topological states. Hence, we propose a series of 2D material candidates with six LGs based on first-principles calculations. Our letter, *for the first time*, offers a method for identifying Dirac points in 2D phononic systems and proposes a series of 2D candidate materials with Dirac phonons at HSPs and on HSLs.

*Symmetry analysis.* In general, the Dirac points in LGs are caused by the four-dimensional (4D) co-representation of little groups in the Brillouin zone (BZ). Recently, Zhang *et al.* [67] proposed the entire co-representation of 528 magnetic LGs by restricting specific co-representations in 3D magnetic space groups. Herein, we present some necessary conditions for the Dirac points in LGs. First, the order of the little co-group of LGs must be greater than or equal to 4 for Dirac points at HSPs. This can be understood using the property of characters of projective representations as follows:

$$\sum_i \chi_i^2(e) = |\mathcal{L}|, \tag{1}$$

where $\chi_i(e)$ is the character of an identity element and is equal to the dimension of irreducible representation, and $\mathcal{L}$ is the little co-group. Notably, the dimension of irreducible representations can be greater than or equal to 2 only for $|\mathcal{L}| \geq 4$. Second, antiunitary elements are necessary for the magnetic little co-group of HSPs. Antiunitary operators play an important role in "sticking" two 2D irreducible representations together because there is no 4D irreducible representation in LGs. The Dirac point on HSLs is formed by two accidentally intersecting bands. Therefore, each band must be doubly degenerate and belong to different co-representations. The above symmetry analysis is consistent with Table I.

**Table I. LG candidates (and their corresponding space groups [SGs]) that can host Dirac points at HSPs or on HSLs in 2D phononic systems. This table also includes the locations of the Dirac phonons, the correspondence generators and labels associated with the Dirac points, and the 2D material candidates.**

| Dirac phonons at HSPs | | | | | | | |
|---|---|---|---|---|---|---|---|
| LG No. | LG symbol | SG No. | SG symbol | Generator | Location | Label | Material |
| 33 | pb2$_1$a | 29 | Pca2$_1$ | $C_{2y}, \sigma_z, \mathcal{T}$ | S | $T_1T_1$ | Ni$_8$As$_4$Si$_4$ |
| 43 | pbaa | 54 | Pcca | $C_{2x}, C_{2z}, I, \mathcal{T}$ | S | $U_1U_2$ | C$_8$H$_8$ |
| 45 | pbma | 57 | Pbcm | $C_{2x}, C_{2y}, I, \mathcal{T}$ | S | $S_1S_2$ | Pb$_4$O$_4$ |
| Dirac phonons on HSLs | | | | | | | |
| LG No. | LG symbol | SG No. | SG symbol | Generator | Location | Label | Material |
| 29 | pb2$_1$m | 26 | Pmc2$_1$ | $\sigma_z, C_{2y}\mathcal{T}$ | Y–S path | $\{B_1B_1, B_2B_2\}$ | Mo$_2$Br$_4$O$_4$ |
| 33 | pb2$_1$a | 29 | Pca2$_1$ | $\sigma_z, C_{2y}\mathcal{T}$ | Y–S path | $\{B_1B_1, B_2B_2\}$ | Ni$_8$As$_4$Si$_4$ |
| 40 | pmam | 51 | Pmma | $C_{2y}, \sigma_z, C_{2z}\mathcal{T}$ | X–S path | $\{A_1A_2, A_3A_4\}$ | |
| 43 | pbaa | 54 | Pcca | $C_{2x}, \sigma_z, C_{2z}\mathcal{T}$ | Y–S path | $\{A_1A_2, A_3A_4\}$ | C$_8$H$_8$ |
| 44 | pbam | 55 | Pbam | $C_{2x}, \sigma_y, C_{2y}\mathcal{T}$ | X–S path | $\{D_1D_2, D_3D_4\}$ | Nb$_4$Te$_8$Si$_2$ |
| | | | | | Y–S path | | |
| 45 | pbma | 57 | Pbcm | $C_{2x}, \sigma_y, C_{2y}\mathcal{T}$ | Y–S path | $\{C_1C_2, C_3C_4\}$ | Pb$_4$O$_4$ |
| 63 | p4/mbm | 127 | P4/mbm | $C_{2x}, \sigma_y, C_{2y}\mathcal{T}$ | X–M path | $\{Y_1Y_2, Y_3Y_4\}$ | Mg$_2$N$_4$ |

*Material candidates with Dirac phonons at HSPs.* Table I shows that LGs. 33, 43, and 45 should exhibit Dirac points, known as fourfold essential degenerate points, at the S HSP. We supported our symmetry analysis by predicting a series of 2D materials with LGs 33, 43, and 45 and multiple Dirac phonons at the S HSP based on first-principles calculations. The structural models for 2D $Ni_8As_4Si_4$, 2D $C_8H_8$, and 2D $Pb_4O_4$ materials are shown in Fig. 1(a)–(c), respectively. Note that the structural models for 2D $Ni_8As_4Si_4$ and 2D $Pb_4O_4$ were screened from the Computational 2D Materials Database [68] and 2D Materials Encyclopedia [69], respectively. He *et al.* [70] proposed the structural model for 2D $C_8H_8$ in 2012.

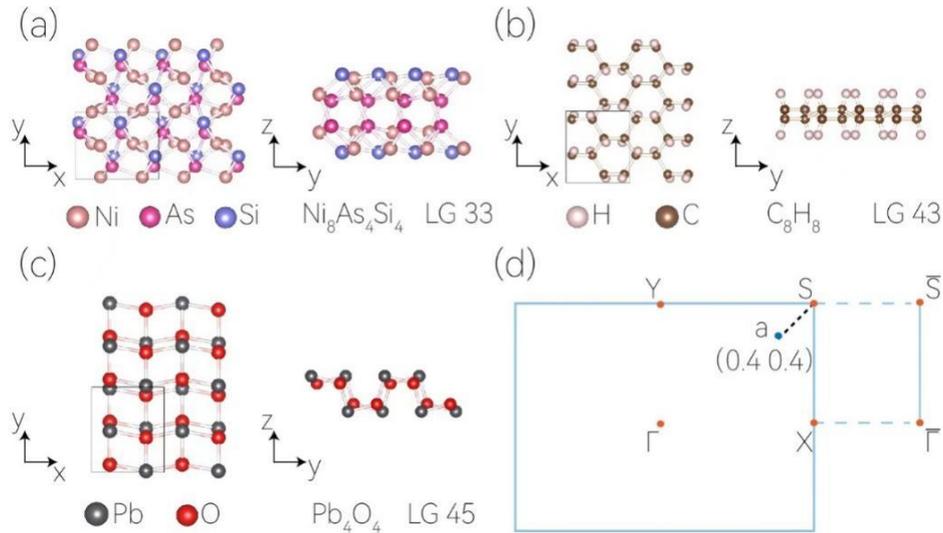

**FIG. 1. (a)–(c) Different side views of the structural models for 2D $Ni_8As_4Si_4$, 2D $C_8H_8$, and 2D $Pb_4O_4$. (d) The 2D BZ and its projection to**

the [100] edge.

The phonon dispersions of the 2D $Ni_8As_4Si_4$, 2D $C_8H_8$, and 2D $Pb_4O_4$ materials along the Γ–X–S–Y–Γ–S high-symmetry paths (see Fig. 1(d)) are shown in Fig. 2, Fig. S1, and Fig. S2 (see the Supplemental Material (SM) [71]), respectively. More details about the computational methods can be found in the SM [71]. Notably, multiple Dirac points are observed at the S HSP in the phonon dispersions of the 2D $Ni_8As_4Si_4$, 2D $C_8H_8$, and 2D $Pb_4O_4$ materials, which agree well with the above symmetry analysis. Figure 2(a) shows that there are 12 visible phonon band crossing points with fourfold degeneracies (marked as red dots) at the S HSP. Figure 2(b) shows the enlarged figures of these Dirac phonons (Nos. 1–12).

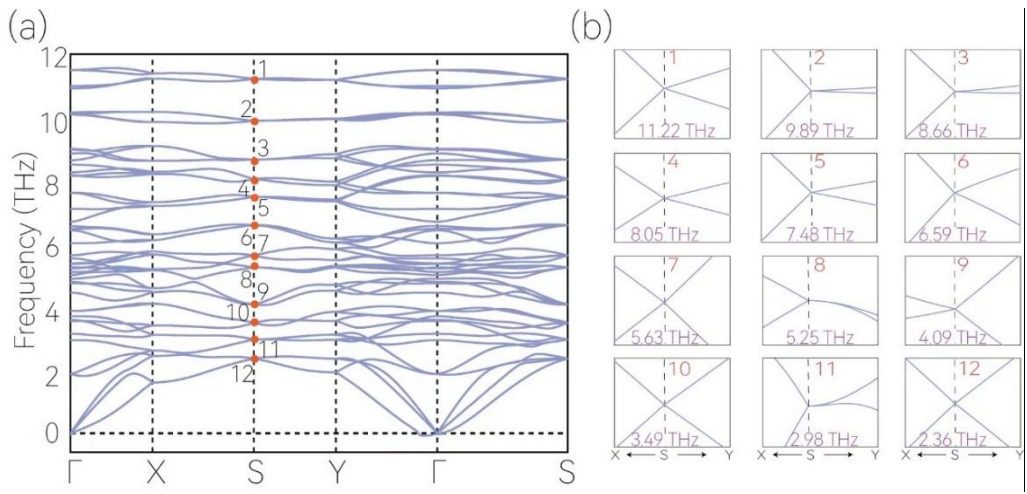

FIG. 2. (a) Phonon dispersions of the 2D 3 × 3 × 1 $Ni_8As_4Si_4$ supercell along the Γ–X–S–Y–Γ–S paths. The Dirac points (labeled with Nos. 1–

**12) located at different frequencies are marked by red dots. (b) Enlarged phonon bands around the 12 Dirac phonons.**

Actually, all phonon bands along the Y–S and X–S high-symmetry paths in 2D materials with LGs 33, 43, and 45 are twofold degenerate. That is, symmetry-enforced twofold degenerate Weyl lines (WLs) should appear along the Y–S and X–S high-symmetry paths in LGs 33, 43, and 45 (see Table SI). Figure 3(a) shows the enlarged phonon bands around the Dirac point (No. 1) at about 11.22 THz as a typical example. The two WLs that form the Dirac point at the S HSP are highlighted by two colors. We selected some symmetry paths, such as b–n–$b_1$, c–m–$c_1$, d–p–$d_1$, and e–o–$e_1$ (see Fig. 3(b)), and calculated the phonon dispersion along them to confirm that all the phononic crossing points of the two WLs have twofold degeneracies. The m and n were on the X–S high–symmetry paths, while o and p were on the Y–S paths. Figures 3(c) and 3(d) show that the twofold degenerate Weyl points appeared at the o, p, m, and n symmetry points.

In contrast to 3D systems that have a twofold screw rotation and time-reversal symmetry $\mathcal{T}$ that must lead to a nodal surface at the $k_\perp = \pi$ plane, where $k_\perp$ is perpendicular to the screw rotation axis, 2D systems have symmetries (*i.e.*, $S_{2x}\mathcal{T}$ and $S_{2y}\mathcal{T}$) that lack a degree of freedom, resulting in different degeneracies. That is, the WL only inherits the degeneracy on one line of the nodal surface because the BZ is 2D.

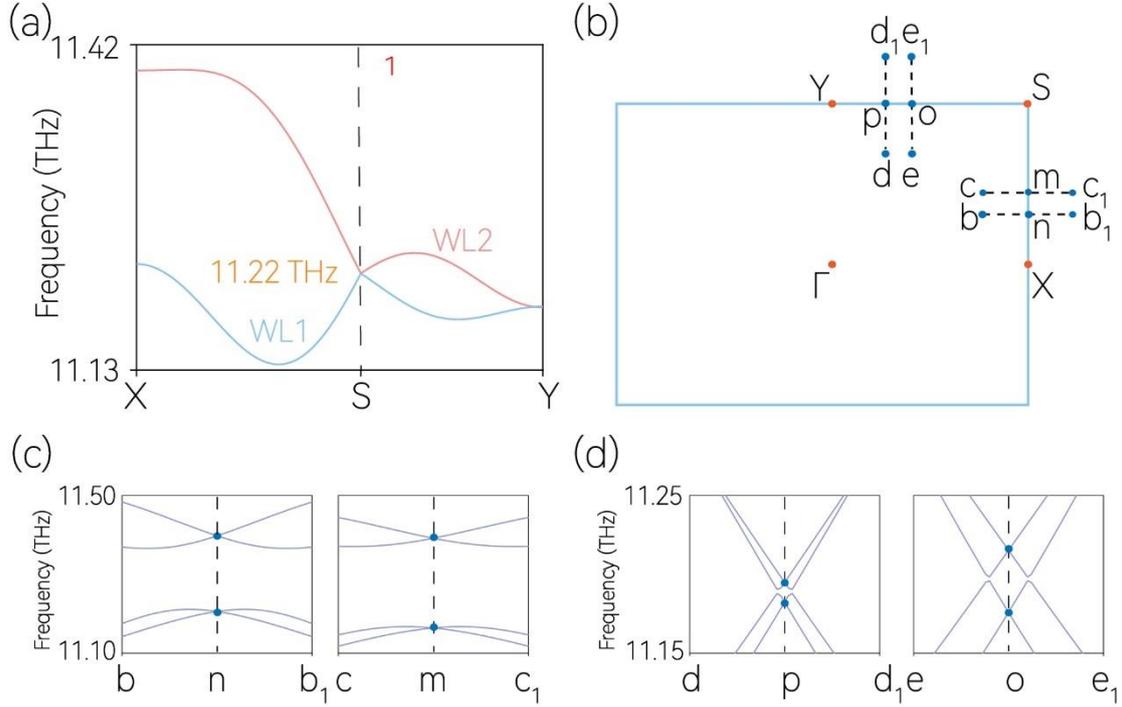

**FIG. 3. (a) Enlarged phonon band of the No. 1 Dirac phonon. Two WLs (red and blue colors) appear along the X–S–Y paths. (b) The 2D BZ and the b–n–$b_1$, c–m–$c_1$, d–p–$d_1$, and e–o–$e_1$ symmetry paths. (c and d) Calculated phonon dispersions along the abovementioned symmetry paths. The Weyl points in (b)–(d) are indicated by blue dots.**

The twofold degenerate WLs along the Y–S and X–S high-symmetry paths intersected at the S HSP and formed a fourfold degenerate point at the S point in LGs 33, 43, and 45. The phonon dispersions around the Dirac points (Nos. 2 and 12) at the S point along the a–S–a paths are shown in

Figs. 4(a) and 4(b), respectively, to verify this. The figures show that the Dirac points have fourfold degeneracies.

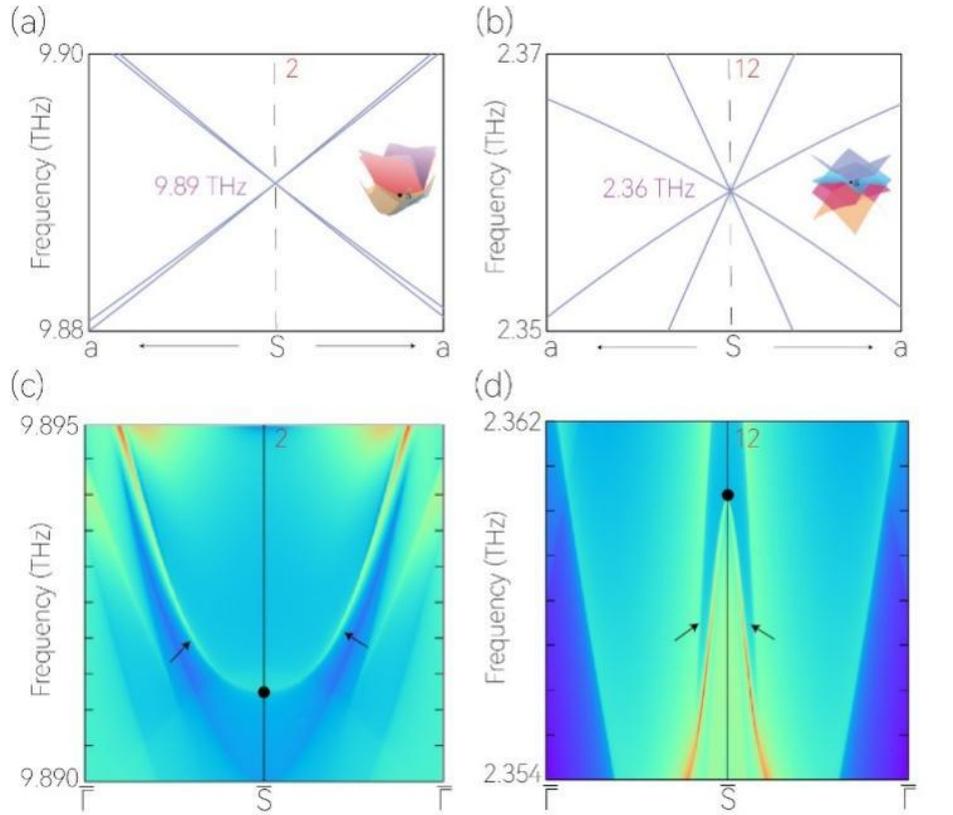

**FIG. 4. (a)–(b) Enlarged phonon dispersions around the Nos. 2 and 12 Dirac phonons, respectively. The insets show the 3D plot of the fourfold degenerate Dirac points (Nos. 2 and 12) at the S HSP. (c and d) Edge states of the Nos. 2 and 12 Dirac phonons, respectively. The black arrows in (c)–(d) show the edge states originating from the projections of the Nos. 2 and 12 Dirac points.**

Furthermore, we studied the edge states of the Dirac phonons (with Nos. 2

and 12) along [100] (see Fig. 1(d)). The results are shown in Figs. 4(c) and 4(d), respectively. The positions of the Dirac points (Nos. 2 and 12) are marked by black balls. The edge states noticeably stem from the projection of the Dirac points. Such clean edge states will benefit follow-up experimental detections using surface-sensitive probes, such as electron energy loss spectroscopy and helium scattering.

*Material candidates with Dirac phonons on HSLs.* Table I shows that LGs 29, 33, 40, 43, 44, 45, and 63 may host Dirac points on HSLs. Note that the Dirac point along HSLs has an accidental fourfold degeneracy. To support our symmetry analysis, six 2D material candidates with LGs 29, 33, 43, 44, 45, and 63, including $Mo_2Br_4O_4$, $Ni_8As_4Si_4$, $C_8H_8$, $Nb_4Te_8Si_2$, $Pb_4O_4$, and $Mg_2N_4$, are presented in this letter. Figures S3–S8 display the structural models, 2D BZ, and calculated phonon dispersions for these six materials (see the SM [71]).

The 2D $Mo_2Br_4O_4$, 2D $Ni_8As_4Si_4$, 2D $C_8H_8$, and 2D $Pb_4O_4$ materials with LGs 29, 33, 43, and 45, respectively, host Dirac points on the Y–S path (see Figs. S3(d), S4(d), S5(d), and S7(d) in the SM [71]). For example, the bands along the Y–S path in 2D $C_8H_8$, which has LG 43, are twofold degenerate (see Table SI in the SM [71]), and two WLs cross each other to form a fourfold degenerate Dirac point around 24 THz (see Fig. S6(d) in the SM [71]). Furthermore, the 2D Nb$_4$Te$_8$Si$_2$ material with LG 44 can host Dirac points on both the X–S and Y–S paths (see Fig. S6 (d) in the SM [71]). The 2D Mg$_2$N$_4$ material with LG 63 can host Dirac points on the X–M path (see Fig. S8 (d) in the SM [71]).

*Spinless lattice models with LG 40.* Unfortunately, 2D materials with LG 40 were not discovered in this study. We constructed a tight-binding (TB) model to demonstrate that Dirac points exist on the X–S path in 2D materials with LG 40 in order to aid further investigations.

We chose a representation of $u = \left(u_{A_x}, u_{A_y}, u_{B_x}, u_{B_y}\right)^T$, where A = (0,0) and B = (1/2,0), which correspond to the displacements of the 2$a$ Wyckoff position along the $x$ and $z$ directions. Thereafter, the four-band dynamic matrix $D(\mathbf{k})$ is satisfied (see Fig. 5(a)):

$$\begin{pmatrix} D_{11} & 0 & D_{13} & 0 \\ & D_{22} & 0 & D_{24} \\ & & D_{11} & 0 \\ \dagger & & & D_{22} \end{pmatrix}, \qquad (2)$$

where $D_{11} = e_1 + 2t_1 \cos k_y$, $D_{13} = 2\cos\frac{k_x}{2} t_3$, $D_{22} = e_2 + 2t_2 \cos k_y$, and $D_{24} = 2\cos\frac{k_x}{2} t_4$. The phonon frequency ($\omega$) can be solved using $D(\mathbf{k})u_\mathbf{k} = \omega^2 u_\mathbf{k}$. Figure 5(b) depicts the phonon dispersion of the spinless lattice model [Eq. (2)] along the Γ–X–S–Y–Γ paths. We set t$_1$ = −0.146, t$_2$ = 0.146, t$_3$ = −0.074, and t$_4$ = −0.07 for the bands in Fig. 5(b). Figures 5(c) and 5(d) show that a fourfold degenerate Dirac point appeared on the X–S path and was formed by the crossing of two WLs.

This simple spinless model may serve as a starting point for future research into the Dirac points in LG 40.

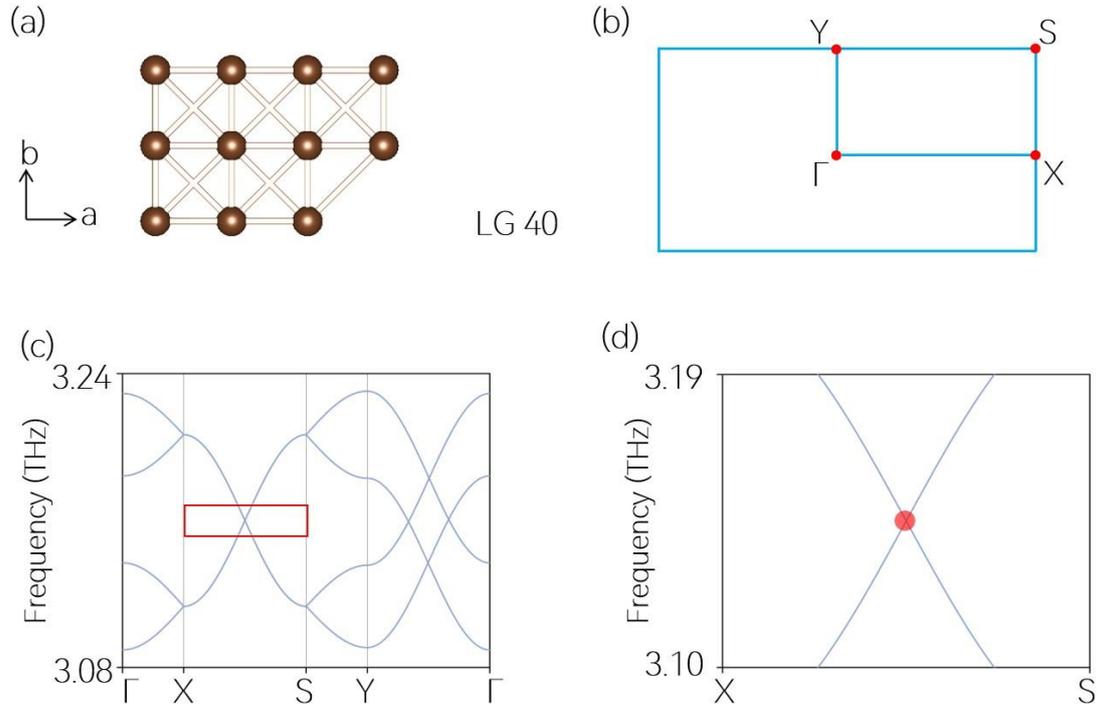

FIG. 5. (a) Spinless TB model with LG 40. (b) The 2D BZ and selected high-symmetry paths. (c) Calculated phonon dispersion for the spinless lattice model with LG 40. The Dirac point location on the X–S path is highlighted by a red box. (d) Enlarged phonon dispersion around the Dirac point (marked by a red dot).

In conclusion, we studied the symmetry conditions of 80 LGs and discovered that Dirac phonons can appear in 7 of the 80 LGs. Specifically, Dirac phonons at the S HSP can only appear in LGs 33, 43, and 45, whereas Dirac phonons on HSLs can appear in LGs 29, 33, 40, 43, 44, 45, and 63.

Thereafter, we predicted several 2D materials that host fourfold degenerate Dirac phonons at HSPs and on HSLs. Note that Chen *et al.* [49] reported the existence of fourfold degenerate Dirac points on HSLs or at HSPs in 3D phononic systems in 2021. However, no other researcher has explored the fourfold degenerate Dirac points in 2D phononic systems. Hence, this letter provides a guideline for studying the fourfold degenerate Dirac points in 2D phononic systems *for the first time*. More importantly, this letter is the first to contribute to the material realization of fourfold degenerate Dirac points in 2D phononic systems. *Undoubtedly*, our letter can be viewed as a guide for investigating not only fourfold degenerate Dirac points in 2D phononic systems but also other types of emergent particles in 2D phononic systems.

References

1. Stern, A., & Lindner, N. H. (2013). Topological quantum computation—from basic concepts to first experiments. *Science*, *339*(6124), 1179-1184.

2. Haldane, F. D. M. (2017). Nobel lecture: Topological quantum matter. *Reviews of Modern Physics*, *89*(4), 040502.

3. Hao, N., & Hu, J. (2019). Topological quantum states of matter in iron-based superconductors: from concept to material realization. *National Science Review*, *6*(2), 213-226.

4. Yu, Z. M., Zhang, Z., Liu, G. B., Wu, W., Li, X. P., Zhang, R. W., ... & Yao, Y. (2022). Encyclopedia of emergent particles in three-dimensional crystals. *Science Bulletin*, *67*(4), 375-380.

5. Wu, W., Liu, Y., Li, S., Zhong, C., Yu, Z. M., Sheng, X. L., ... & Yang, S. A. (2018). Nodal surface semimetals: Theory and material realization. *Physical Review B*, *97*(11), 115125.

6. Wang, K., Dai, J. X., Shao, L. B., Yang, S. A., & Zhao, Y. X. (2020). Boundary criticality of PT-invariant topology and second-order nodal-line semimetals. *Physical Review Letters*, *125*(12), 126403.

7. Cao, L., Zhou, G., Wu, Q., Yang, S. A., Yang, H. Y., Ang, Y. S., & Ang, L. K. (2020). Electrical contact between an ultrathin topological Dirac semimetal and a two-dimensional material. *Physical Review Applied*, *13*(5), 054030.

8. Chen, C., Zeng, X. T., Chen, Z., Zhao, Y. X., Sheng, X. L., & Yang, S. A. (2022). Second-Order Real Nodal-Line Semimetal in Three-Dimensional Graphdiyne. *Physical Review Letters*, *128*(2), 026405.

9. Wan, Q., Yang, T. Y., Li, S., Yang, M., Zhu, Z., Wu, C. L., ... & Xu, N. (2021). Inherited weak topological insulator signatures in the topological hourglass semimetal Nb 3 X Te 6 (X= Si, Ge). *Physical Review B*, *103*(16), 165107.

10. Zhu, Z., Liu, Y., Yu, Z. M., Wang, S. S., Zhao, Y. X., Feng, Y., ... & Yang, S. A. (2018). Quadratic contact point semimetal: Theory and material realization. *Physical Review B*, *98*(12), 125104.

11. Yang, S. A., Pan, H., & Zhang, F. (2014). Dirac and Weyl superconductors in three dimensions. *Physical review letters*, *113*(4), 046401.

12. Li, S., Liu, Y., Fu, B., Yu, Z. M., Yang, S. A., & Yao, Y. (2018). Almost ideal nodal-loop semimetal in monoclinic CuTeO 3 material. *Physical Review B*, *97*(24), 245148.

13. You, J. Y., Chen, C., Zhang, Z., Sheng, X. L., Yang, S. A., & Su, G. (2019). Two-dimensional Weyl half-semimetal and tunable quantum anomalous Hall effect. *Physical Review B*, *100*(6), 064408.

14. Zhang, X., Yu, Z. M., Zhu, Z., Wu, W., Wang, S. S., Sheng, X. L., & Yang, S. A. (2018). Nodal loop and nodal surface states in the Ti 3 Al family of materials. *Physical Review B*, *97*(23), 235150.

15. Zhu, Z., Yu, Z. M., Wu, W., Zhang, L., Zhang, W., Zhang, F., & Yang, S. A. (2019). Composite Dirac semimetals. *Physical Review B*, *100*(16), 161401.

16. Burkov, A. A. (2016). Topological semimetals. *Nature materials*, *15*(11), 1145-1148.

17. Lv, B. Q., Qian, T., & Ding, H. (2021). Experimental perspective on three-dimensional topological semimetals. *Reviews of Modern Physics*, *93*(2), 025002.

18. Fang, C., Lu, L., Liu, J., & Fu, L. (2016). Topological semimetals with helicoid surface states. *Nature Physics*, *12*(10), 936-941.

19. Fang, C., Gilbert, M. J., Dai, X., & Bernevig, B. A. (2012). Multi-Weyl topological semimetals stabilized by point group symmetry. *Physical review letters*, *108*(26), 266802.

20. Chen, J., Li, H., Ding, B., Zhang, H., Liu, E., & Wang, W. (2021). Large anomalous Hall angle in a topological semimetal candidate TbPtBi. *Applied Physics Letters*, *118*(3), 031901.

21. Sun, X. C., He, C., Liu, X. P., Lu, M. H., Zhu, S. N., & Chen, Y. F. (2017). Two-dimensional topological photonic systems. *Progress in Quantum Electronics*, *55*, 52-73.

22. Khanikaev, A. B., & Shvets, G. (2017). Two-dimensional topological photonics. *Nature photonics*, *11*(12), 763-773.

23. Ozawa, T., Price, H. M., Amo, A., Goldman, N., Hafezi, M., Lu, L., ...

& Carusotto, I. (2019). Topological photonics. *Reviews of Modern Physics*, *91*(1), 015006.

24. Lu, L., Joannopoulos, J. D., & Soljačić, M. (2014). Topological photonics. *Nature photonics*, *8*(11), 821-829.

25. Arregui, G., Ortiz, O., Esmann, M., Sotomayor-Torres, C. M., Gomez-Carbonell, C., Mauguin, O., ... & Lanzillotti-Kimura, N. D. (2019). Coherent generation and detection of acoustic phonons in topological nanocavities. *APL Photonics*, *4*(3), 030805.

26. Park, S., Hwang, Y., Choi, H. C., & Yang, B. J. (2021). Topological acoustic triple point. *Nature communications*, *12*(1), 1-9.

27. Ding, Z. K., Zeng, Y. J., Pan, H., Luo, N., Zeng, J., Tang, L. M., & Chen, K. Q. (2022). Edge states of topological acoustic phonons in graphene zigzag nanoribbons. *Physical Review B*, *106*(12), L121401.

28. Yang, Z., Gao, F., Shi, X., Lin, X., Gao, Z., Chong, Y., & Zhang, B. (2015). Topological acoustics. *Physical review letters*, *114*(11), 114301.

29. Zhou, P., Liu, G. G., Yang, Y., Hu, Y. H., Ma, S., Xue, H., ... & Zhang, B. (2020). Observation of photonic antichiral edge states. *Physical Review Letters*, *125*(26), 263603.

30. Liu, G. G., Yang, Y., Ren, X., Xue, H., Lin, X., Hu, Y. H., ... & Zhang, B. (2020). Topological Anderson insulator in disordered photonic crystals. *Physical Review Letters*, *125*(13), 133603.

31. Liu, Y., Chen, X., & Xu, Y. (2020). Topological phononics: from

fundamental models to real materials. *Advanced Functional Materials*, *30*(8), 1904784.

32. Liu, Y., Xu, Y., & Duan, W. (2019). Three-dimensional topological states of phonons with tunable pseudospin physics. *Research*, *2019*.

33. Liu, Y., Xu, Y., & Duan, W. (2018). Berry phase and topological effects of phonons. *National Science Review*, *5*(3), 314-316.

34. Liu, Y., Zou, N., Zhao, S., Chen, X., Xu, Y., & Duan, W. (2022). Ubiquitous topological states of phonons in solids: Silicon as a model material. *Nano letters*, *22*(5), 2120-2126.

35. Chen, X. Q., Liu, J., & Li, J. (2021). Topological phononic materials: Computation and data. *The Innovation*, *2*(3), 100134.

36. Chen, J., He, J., Pan, D., Wang, X., Yang, N., Zhu, J., ... & Zhang, G. (2022). Emerging theory and phenomena in thermal conduction: A selective review. *Science China Physics, Mechanics & Astronomy*, *65*(11), 1-16.

37. Chen, J., Xu, X., Zhou, J., & Li, B. (2022). Interfacial thermal resistance: Past, present, and future. *Reviews of Modern Physics*, *94*(2), 025002.

38. Li, N., Ren, J., Wang, L., Zhang, G., Hänggi, P., & Li, B. (2012). Colloquium: Phononics: Manipulating heat flow with electronic analogs and beyond. *Reviews of Modern Physics*, *84*(3), 1045.

39. Li, J., Liu, J., Baronett, S. A., Liu, M., Wang, L., Li, R., ... & Chen, X. Q. (2021). Computation and data driven discovery of topological phononic materials. *Nature communications*, *12*(1), 1-12.

40. Zhang, T., Song, Z., Alexandradinata, A., Weng, H., Fang, C., Lu, L., & Fang, Z. (2018). Double-Weyl phonons in transition-metal monosilicides. *Physical review letters*, *120*(1), 016401.

41. Miao, H., Zhang, T. T., Wang, L., Meyers, D., Said, A. H., Wang, Y. L., ... & Dean, M. P. M. (2018). Observation of double Weyl phonons in parity-breaking FeSi. *Physical review letters*, *121*(3), 035302.

42. Liu, Q. B., Wang, Z., & Fu, H. H. (2021). Charge-four Weyl phonons. *Physical Review B*, *103*(16), L161303.

43. Liu, Q. B., Qian, Y., Fu, H. H., & Wang, Z. (2020). Symmetry-enforced Weyl phonons. *npj Computational Materials*, *6*(1), 1-6.

44. Wang, R., Xia, B. W., Chen, Z. J., Zheng, B. B., Zhao, Y. J., & Xu, H. (2020). Symmetry-protected topological triangular Weyl complex. *Physical Review Letters*, *124*(10), 105303.

45. Jin, Y. J., Chen, Z. J., Xiao, X. L., & Xu, H. (2021). Tunable double Weyl phonons driven by chiral point group symmetry. *Physical Review B*, *103*(10), 104101.

46. Ding, G., Zhou, F., Zhang, Z., Yu, Z. M., & Wang, X. (2022). Charge-two Weyl phonons with type-III dispersion. *Physical Review B*, *105*(13), 134303.

47. Zhong, M., Han, Y., Wang, J., Liu, Y., Wang, X., & Zhang, G. (2022). Material realization of double-Weyl phonons and phononic double-

helicoid surface arcs with P 2 1 3 space group. *Physical Review Materials*, *6*(8), 084201.

48. Huang, Z., Chen, Z., Zheng, B., & Xu, H. (2020). Three-terminal Weyl complex with double surface arcs in a cubic lattice. *Npj Computational Materials*, *6*(1), 1-7.

49. Chen, Z. J., Wang, R., Xia, B. W., Zheng, B. B., Jin, Y. J., Zhao, Y. J., & Xu, H. (2021). Three-dimensional Dirac phonons with inversion symmetry. *Physical Review Letters*, *126*(18), 185301.

50. Feng, Y., Xie, C., Chen, H., Liu, Y., & Wang, X. (2022). Dirac point phonons at high-symmetry points: Towards materials realization. *Physical Review B*, *106*(13), 134307.

51. Singh, S., Wu, Q., Yue, C., Romero, A. H., & Soluyanov, A. A. (2018). Topological phonons and thermoelectricity in triple-point metals. *Physical Review Materials*, *2*(11), 114204.

52. Sreeparvathy, P. C., Mondal, C., Barman, C. K., & Alam, A. (2022). Coexistence of multifold and multidimensional topological phonons in KMgBO 3. *Physical Review B*, *106*(8), 085102.

53. You, J. Y., Sheng, X. L., & Su, G. (2021). Topological gimbal phonons in t-carbon. *Physical Review B*, *103*(16), 165143.

54. Jin, Y. J., Chen, Z. J., Xia, B. W., Zhao, Y. J., Wang, R., & Xu, H. (2018). Ideal intersecting nodal-ring phonons in bcc C 8. *Physical Review B*, *98*(22), 220103.

55. Xie, C., Liu, Y., Zhang, Z., Zhou, F., Yang, T., Kuang, M., ... & Zhang, G. (2021). Sixfold degenerate nodal-point phonons: Symmetry analysis and materials realization. *Physical Review B*, *104*(4), 045148.

56. Zhou, F., Zhang, Z., Chen, H., Kuang, M., Yang, T., & Wang, X. (2021). Hybrid-type nodal ring phonons and coexistence of higher-order quadratic nodal line phonons in an AgZr alloy. *Physical Review B*, *104*(17), 174108.

57. Zhou, F., Chen, H., Yu, Z. M., Zhang, Z., & Wang, X. (2021). Realistic cesium fluogermanate: An ideal platform to realize the topologically nodal-box and nodal-chain phonons. *Physical Review B*, *104*(21), 214310.

58. Liu, G., Jin, Y., Chen, Z., & Xu, H. (2021). Symmetry-enforced straight nodal-line phonons. *Physical Review B*, *104*(2), 024304.

59. Xie, C., Yuan, H., Liu, Y., Wang, X., & Zhang, G. (2021). Three-nodal surface phonons in solid-state materials: Theory and material realization. *Physical Review B*, *104*(13), 134303.

60. Peng, B., Murakami, S., Monserrat, B., & Zhang, T. (2021). Degenerate topological line surface phonons in quasi-1D double helix crystal SnIP. *npj Computational Materials*, *7*(1), 1-8.

61. Yu, W. W., Liu, Y., Tian, L., He, T., Zhang, X., & Liu, G. (2022). Phononic linear and quadratic nodal points in monolayer XH (X= Si, Ge, Sn). *Journal of Physics: Condensed Matter*, *34*(15), 155703.

62. Yu, W. W., Liu, Y., Meng, W., Liu, H., Gao, J., Zhang, X., & Liu, G. (2022). Phononic higher-order nodal point in two dimensions. *Physical Review B*, *105*(3), 035429.

63. Li, J., Wang, L., Liu, J., Li, R., Zhang, Z., & Chen, X. Q. (2020). Topological phonons in graphene. *Physical Review B*, *101*(8), 081403.

64. Jin, Y., Wang, R., & Xu, H. (2018). Recipe for Dirac phonon states with a quantized valley berry phase in two-dimensional hexagonal lattices. *Nano letters*, *18*(12), 7755-7760.

65. Yang, S. A. (2016, June). Dirac and Weyl materials: fundamental aspects and some spintronics applications. In *Spin* (Vol. 6, No. 02, p. 1640003). World Scientific Publishing Company.

66. Feng, X., Zhu, J., Wu, W., & Yang, S. A. (2021). Two-dimensional topological semimetals. *Chinese Physics B*, *30*(10), 107304.

67. Zhang, Z., Wu, W., Liu, G. B., Yu, Z. M., Yang, S. A., & Yao, Y. (2022). Encyclopedia of emergent particles in 528 magnetic layer groups and 394 magnetic rod groups. *arXiv preprint arXiv:2210.11080*.

68. https://cmrdb.fysik.dtu.dk/c2db/row/As4Si4Ni8-0c60c11fbd24

69. http://www.2dmatpedia.org/2dmaterials/doc/2dm-3669

70. He, C., Zhang, C. X., Sun, L. Z., Jiao, N., Zhang, K. W., & Zhong, J. (2012). Structure, stability and electronic properties of tricycle type graphane. *physica status solidi (RRL)–Rapid Research Letters*, *6*(11), 427-429.

71. See Supplemental Material [SM] for the computational methods, the candidate LGs that can host WL along HSLs in 2D phononic systems, and the structural models, 2D BZ, and the calculated phonon dispersions for $Mo_2Br_4O_4$, $Ni_8As_4Si_4$, $C_8H_8$, $Nb_4Te_8Si_2$, $Pb_4O_4$, and $Mg_2N_4$, with LGs 29, 33, 43, 44, 45, 63., which includes references [72-76].

72. J. Hafner, J. Comput. Chem. 29, 2044 (2008).

73. P. E. Blöchl, Phys. Rev. B 50, 17953 (1994).

74. J. P. Perdew, K. Burke, and M. Ernzerhof, Phys. Rev. Lett. 77, 3865 (1996).

75. A. Togo and I. Tanaka, Scr. Mater. 108, 1 (2015).

76. Q. Wu, S. Zhang, H.-F. Song, M. Troyer, and A. A. Soluyanov, Phys. Chem. 430 431 Commun. 224, 405 (2018).

**Computational Methods**

First-principles calculations were carried out using the Vienna ab initio Simulation Package (VASP) [S1]. With a kinetic energy cutoff of 500 eV, the projector augmented-wave (PAW) [S2] method was used for the plane-wave basis. The exchange–correlation interactions were described by the Perdew–Burke–Ernzerhof (PBE) functional within the generalized gradient approximation (GGA) [S3]. During structural optimizations, the lattice constants and atomic positions were fully relaxed until the total energy and atomic force were less than $10^{-6}$ eV and -0.001 eV Å$^{-1}$, respectively. We used the density functional perturbation theory to obtain the force constants, as implemented in the VASP. Then, we used PHONOPY package [S4] to calculate the phonon dispersion spectrum. A phonon tight-binding Hamiltonian was constructed using the WANNIERTOOLS code [S5]. Green's function iterative method was used to obtain edge states for these 2D materials.

References
S1. J. Hafner, J. Comput. Chem. 29, 2044 (2008).
S2. P. E. Blöchl, Phys. Rev. B 50, 17953 (1994).
S3. J. P. Perdew, K. Burke, and M. Ernzerhof, Phys. Rev. Lett. 77, 3865 (1996).
S4. A. Togo and I. Tanaka, Scr. Mater. 108, 1 (2015).
S5. Q. Wu, S. Zhang, H.-F. Song, M. Troyer, and A. A. Soluyanov, Phys. Chem. 430 431 Commun. 224, 405 (2018).

**Table SI. The candidate LGs (and the corresponding SGs) that can host WL along HSLs in 2D phononic systems. The Location of the WL phonons, the correspondence generators and label associated to the WLs, and the candidate 2D materials are also given in this table.**

| The Weyl nodal line phonons along HSLs | | | | | | | | | |
|---|---|---|---|---|---|---|---|---|---|
| LG No. | LG symbol | Generators | Location | Label | LG No. | LG symbol | Generators | Location | Label |
| 29 | pb2$_1$m | $\sigma_z, C_{2y}\mathcal{T}$ | Y-S | B$_1$B$_1$ B$_2$B$_2$ | 33 | pb2$_1$a | $\sigma_z, C_{2y}\mathcal{T}$ | X-S | B$_1$B$_1$ B$_2$B$_2$ |
| | | | | | | | $C_{2y}, \sigma_z$ | Y-S | H$_1$ |
| 40 | pmam | $C_{2y}, \sigma_z, C_{2z}\mathcal{T}$ | X-S | A$_1$A$_2$ A$_3$A$_4$ | 43 | pbaa | $C_{2y}, \sigma_z, C_{2z}\mathcal{T}$ | X-S | A$_1$ |
| | | | | | | | $C_{2x}, \sigma_y, C_{2y}\mathcal{T}$ | Y-S | G$_1$G$_2$ G$_3$G$_4$ |
| 44 | pbam | $C_{2y}, \sigma_x, C_{2x}\mathcal{T}$ | X-S | G$_1$ | 45 | pbma | $C_{2x}, \sigma_y, C_{2y}\mathcal{T}$ | X-S | D$_1$ |
| | | $C_{2x}, \sigma_z, C_{2z}\mathcal{T}$ | Y-S | A$_1$A$_2$ A$_3$A$_4$ | | | $C_{2x}, \sigma_y, C_{2y}\mathcal{T}$ | Y-S | C$_1$C$_2$ C$_3$C$_4$ |
| 63 | p4/mbm | $C_{2x}, \sigma_y, C_{2y}\mathcal{T}$ | X-M | Y$_1$Y$_2$ Y$_3$Y$_4$ | | | | | |

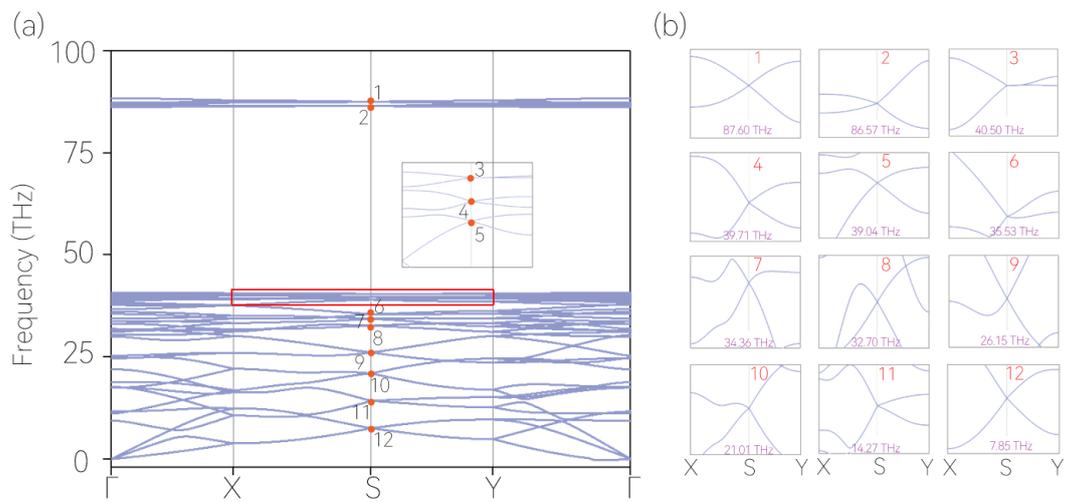

**Figure S1. (a) Phonon dispersion of the 2D 2×2×1 C$_8$H$_8$ supercell along the Γ-X-S-Y-Γ paths. The Dirac points (labeled with Nos. 1-12) located at different frequencies are marked by red dots. (b) Enlarged phonon bands around the 12 Dirac phonons.**

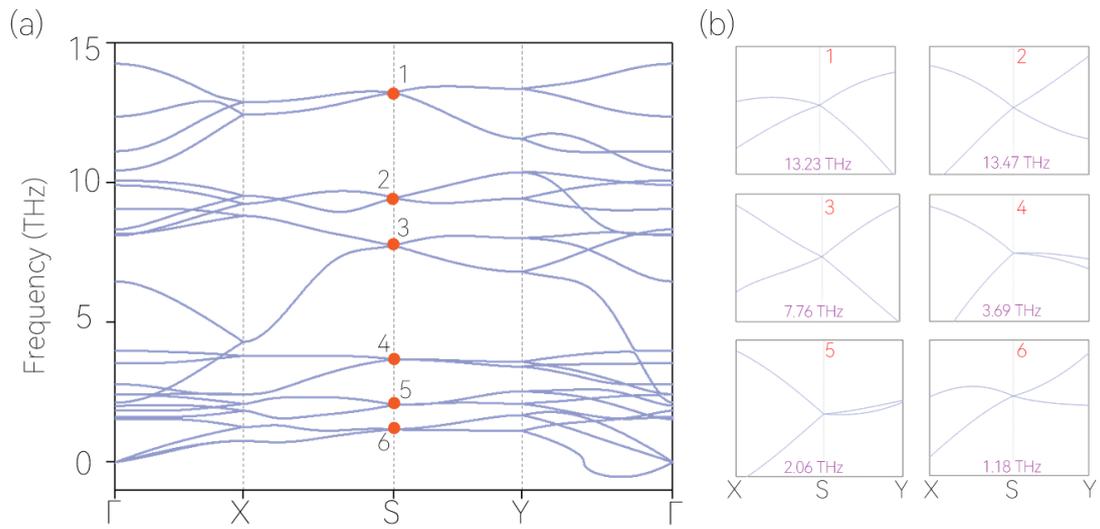

**Figure S2.** (a) Phonon dispersion of the 2D 2×2×1 $Pb_4O_4$ supercell along the Γ-X-S-Y-Γ paths. The Dirac points (labeled with Nos. 1-6) located at different frequencies are marked by red dots. (b) Enlarged phonon bands around the six Dirac phonons.

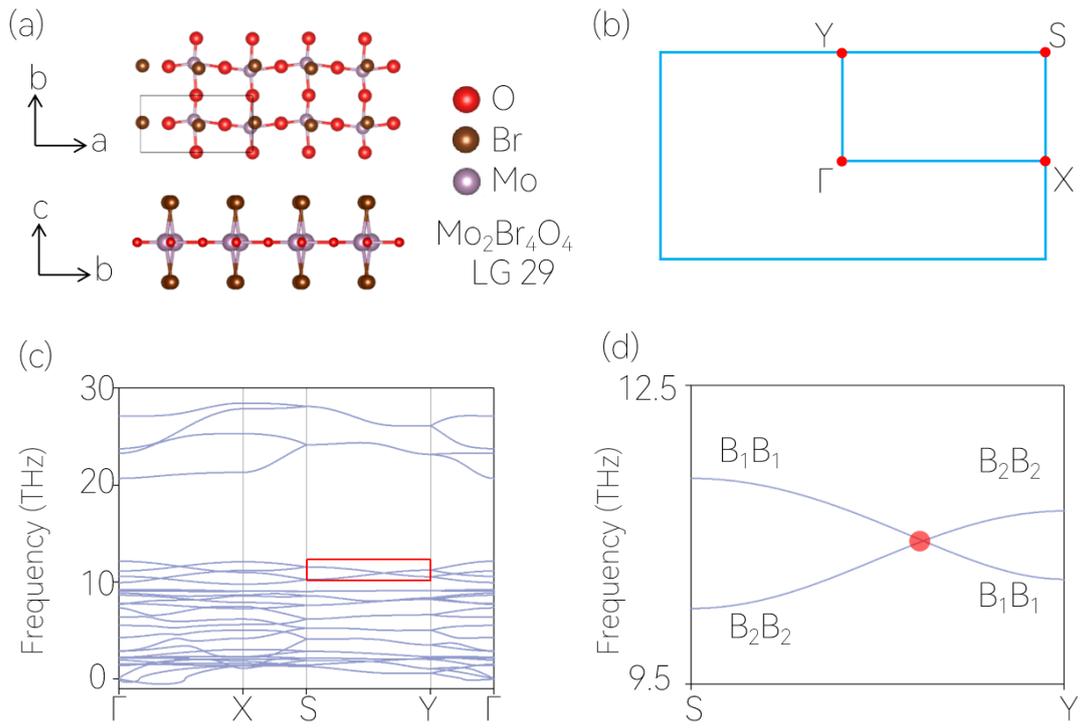

**Figure S3.** (a) Structural model of the 2D 2×2×1 Mo$_2$BrO$_4$ supercell with LG 29. (b) 2D BZ and selected high-symmetry paths. (c) Phonon dispersion of 2D Mo$_2$BrO$_4$ along the Γ-X-S-Y-Γ paths. (d) Enlarged phonon dispersion around the Dirac point on S-Y path. The Dirac point is marked by red dot.

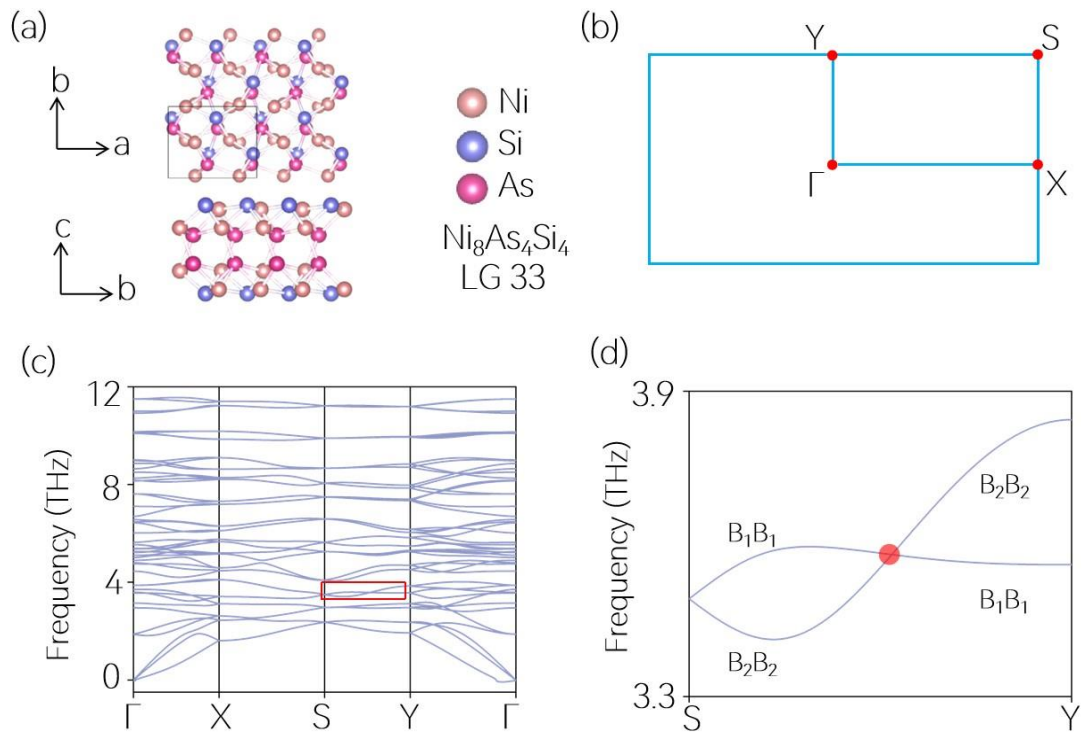

**Figure S4.** (a) Structural model of the 2D 3×3×1 $Ni_8As_4Si_4$ supercell with LG 33. (b) 2D BZ and selected high-symmetry paths. (c) Phonon dispersion of the 2D $Ni_8As_4Si_4$ along the Γ-X-S-Y-Γ paths. (d) Enlarged phonon dispersion around the Dirac point on S-Y path. The Dirac point is marked by red dot.

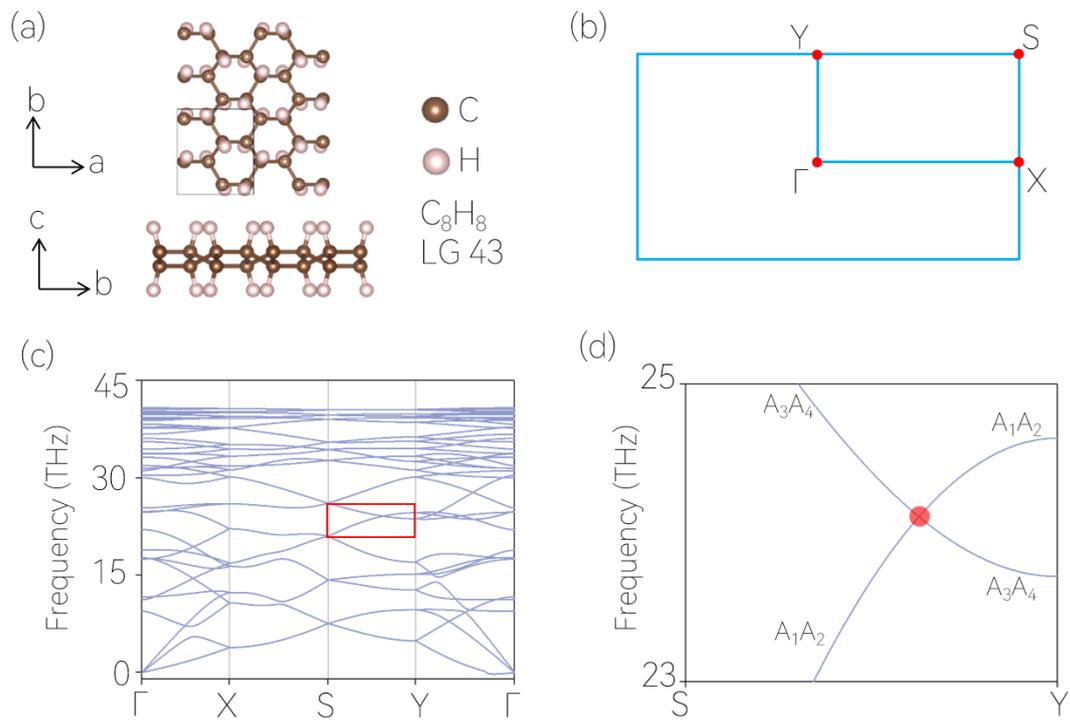

**Figure S5.** (a) Structural model of the 2D 2×2×1 $C_8H_8$ supercell with LG 43. (b) 2D BZ and selected high-symmetry paths. (c) Phonon dispersion of the 2D $C_8H_8$ along the Γ-X-S-Y-Γ paths. (d) Enlarged phonon dispersion around the Dirac point on S-Y path. The Dirac point is marked by red dot.

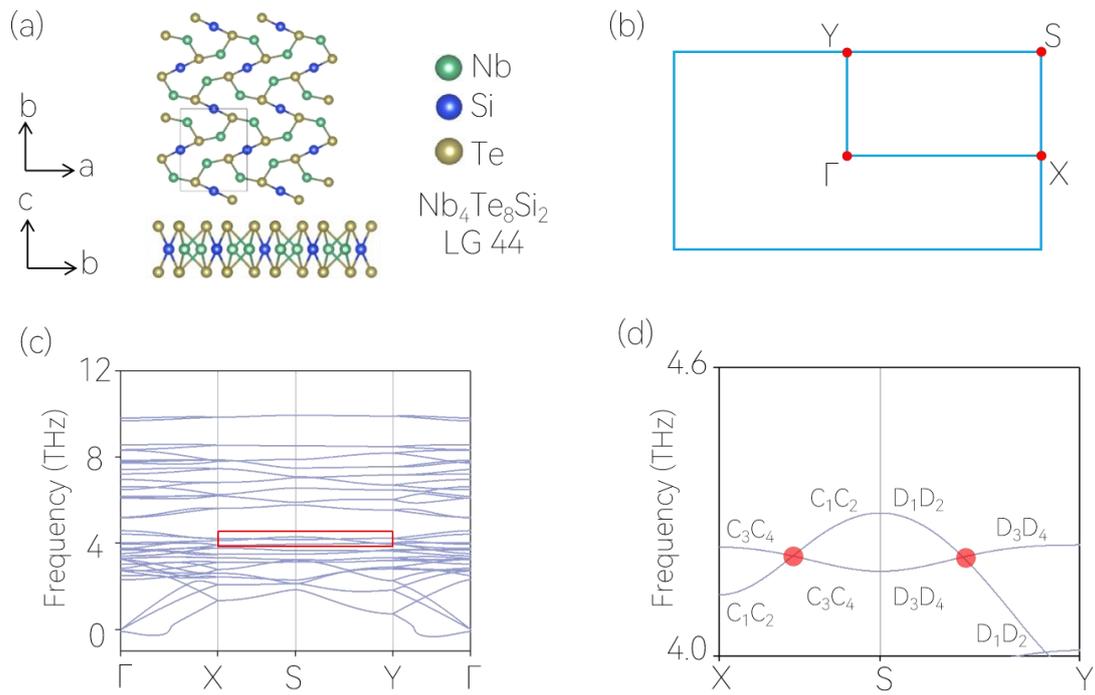

**Figure S6.** (a) Structural model of the 2D 2×2×1 $Nb_4Te_8Si_2$ supercell with LG 44. (b) 2D BZ and selected high-symmetry paths. (c) Phonon dispersion of the 2D $Nb_4Te_8Si_2$ along the Γ-X-S-Y-Γ paths. (d) Enlarged phonon dispersion around the Dirac points on X-S-Y path. The Dirac points are marked by red dots.

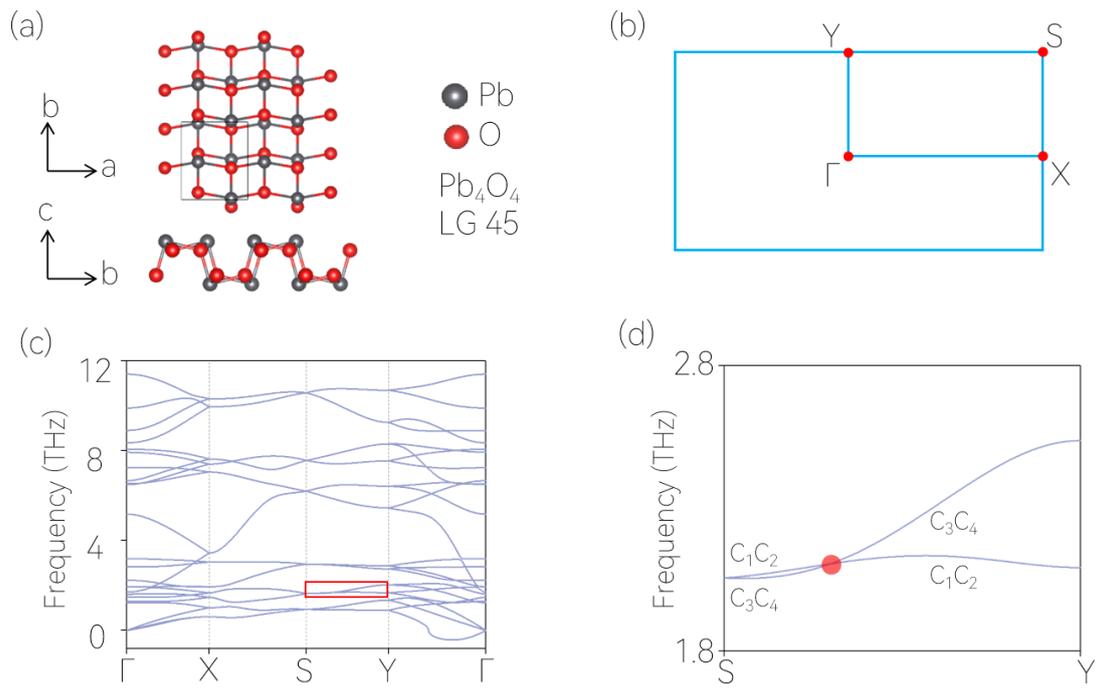

**Figure S7.** (a) Structural model of the 2D 2×2×1 Pb$_4$O$_4$ supercell with LG 45. (b) 2D BZ and selected high-symmetry paths. (c) Phonon dispersion of 2D Pb$_4$O$_4$ along the Γ-X-S-Y-Γ paths. (d) Enlarged phonon dispersion around the Dirac point on S-Y path. The Dirac point is marked by red dot.

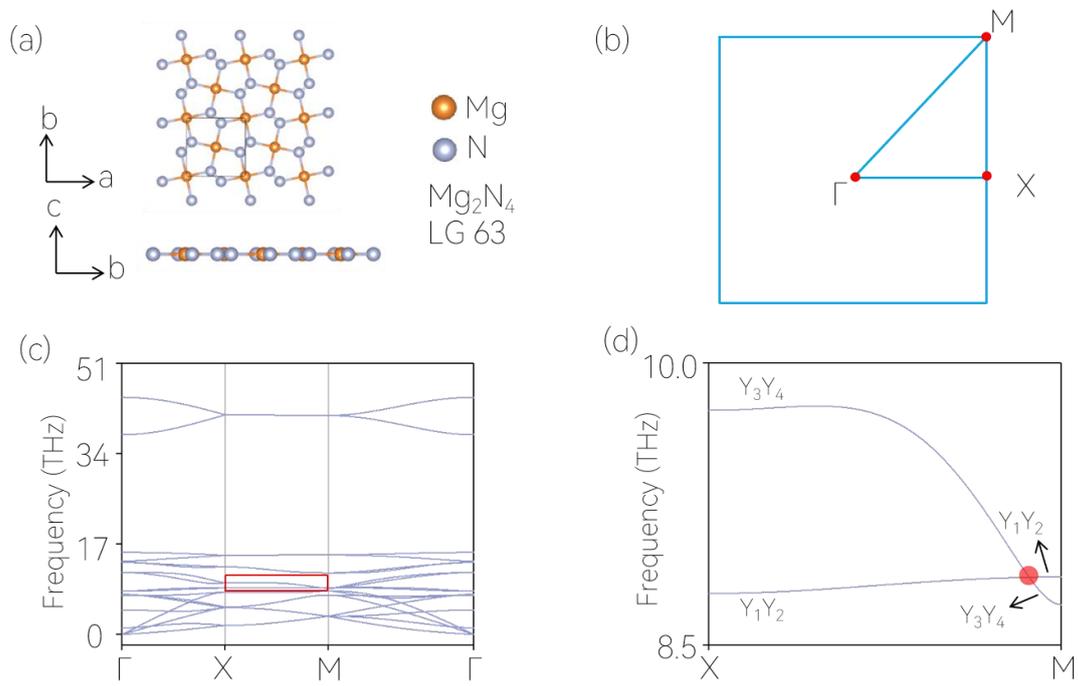

**Figure S8.** (a) Structural model of the 2D 2×2×1 Mg$_2$N$_4$ supercell with LG 63. (b) 2D BZ and selected high-symmetry paths. (c) Phonon dispersion of 2D Mg$_2$N$_4$ along the Γ-X-M-Γ paths. (d) Enlarged phonon dispersion around the Dirac point on X-M path. The Dirac point is marked by red dot.